# Addendum to "Monomer motion in single- and double-stranded DNA coils"
### [arXiv: cond-mat/0509399]

J. Tothova[1], B. Brutovsky[1], V. Lisy[1,2]

[1]*Institute of Physics, P.J. Safarik University, Jesenna 5, 041 54 Kosice, Slovakia*
[2]*Department of Physics, Faculty of Electrical Engineering and Informatics, Technical University in Kosice, Park Komenskeho 2, 040 20 Kosice, Slovakia*

In our work [J. Tothova *et al.*, arXiv:cond-mat/0509399] the first observation of the kinetics of individual polymer monomers using the fluorescence correlation technique [R. Shusterman *et al.*, Phys. Rev. Lett. 92, 048303 (2004)] has been interpreted within the joint Rouse-Zimm theory. Optimizing the theory to the experimental data the phenomenological parameters for the statistical-mechanical description of the universal behavior of double and single stranded DNA and the dominant types of their dynamics have been determined. Recently, these data have been corrected [R. Shusterman *et al.*, Phys. Rev. Lett. 98, 029901 (2007)]. In this Addendum the fits of the theory to the new data are presented. The main conclusions of our preceding work remain unchanged. Moreover, the new data allow a significantly better agreement with the theory than the previous ones.

**1.** In our works [1, 2] the experiments [3] in which the kinetics of individual monomers within the polymer coils was observed for the first time using the fluorescence correlation technique have been interpreted. Our analysis of the experimental data was based on the joint Rouse-Zimm theory that contains the pure Rouse and Zimm models as limiting cases. Applying this more general approach, the phenomenological parameters of polymers that we have determined from the experiments significantly differ from their values used in Ref. [3] where the dynamics of single- and double-stranded DNA (ss and dsDNA) was studied assuming the validity of the Rouse and Zimm limits. The found parameters allowed us to conclude that the kinetics of long dsDNA polymers is mainly of the Zimm type rather than the Rouse one as proposed in Ref. [3]. However, recently [4], the experimental data [3] on the end monomer displacement as a function of time have been corrected. The correction amounts to a uniform shift of all the experimental curves in a log-log plot. According to Ref. [4], the correction does not affect the power laws characterizing the observed kinetic regimes and does not change conclusions of the previous work [3].

Using the new data, we have repeated the fits of the theoretical mean square displacement (MSD) of the end monomers of the studied DNA polymers. The method of optimization of the theory to the experimental data was the same as in the preceding paper [1]. Figures 1 and 2 show the examples of the new fits corresponding to those given in Ref. [1] (Figs. 3 and 4). The fits yielded the polymer parameters that differ from the parameters found in Ref. [1]. So, in the case of dsDNA the mean square distance between the beads along the chain is somewhat smaller ($a$ = 84.6 nm instead of 99.1 nm [1]) but the bead radius changed notably ($b$ = 6.3 nm instead of 49.5 nm). Due to this the draining parameter $h$ is approximately 1.3 instead of 8.6 (for details see Ref. [1]). The draining parameter for the diffusion of the whole coil is $D_Z/D_R = 4\sqrt{2}h/3 \approx 2.4$, where $D_Z$ and $D_R$ are the coil diffusion coefficients in the Zimm and Rouse limit, respectively. The Kuhn length for this polymer is $l \approx 71$ nm (98 nm in [1]). We remind that the parameters used in Ref. [3] lead to the unrealistic value $l \approx 30$ nm. For ssDNA (Fig. 2) the found mean square distance between the beads is approximately 8 nm (9 nm in [1]), the bead radius is 4 nm (instead of 4.6 nm), and the number of beads $N = 474$ (instead of 422). Thus the draining parameter has changed only slightly: $h \approx 21$ (20 in [1]). Finally, the Kuhn length $l \approx 7.9$ nm (for the distance between the bases along the chain 0.58 nm [5]), while in Ref. [1] the value 9 nm has been determined.

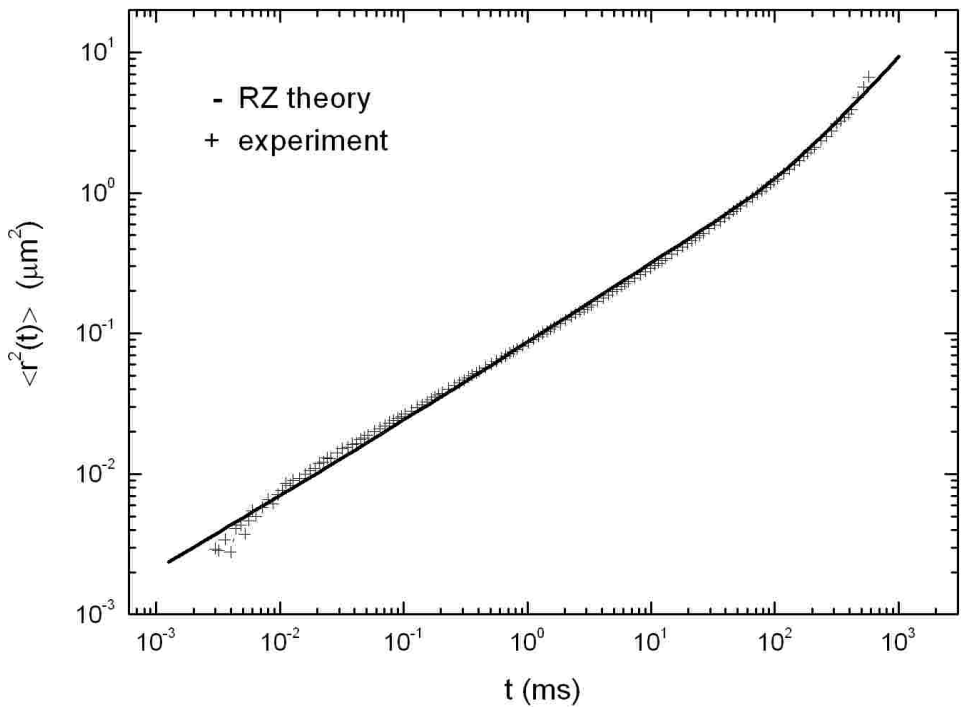

**Fig. 1.** MSD for the joint Rouse-Zimm (RZ) model optimized to the experimental data [4] for dsDNA (23100 bp) in aqueous solution at $T = 293$ K and $\eta = 1$ mPa s. The polymer parameters are $N = 78$, $a = 84.58$ nm, and $b = 6.34$ nm with the draining parameter $h \approx 1.3$.

**2.** Comparing the presented fits to the new experimental data one can see that, as distinct from our previous papers [1, 2], the agreement with the theory is very good. Especially it concerns the description of the dynamics of ssDNA, which is known to be a flexible polymer so that the Rouse-Zimm theory should be well applicable in this case (note that there was a significant discrepancy between the theory and the erroneous experimental data [3]; this discrepancy could not be resolved for any set of the polymer parameters). A large draining parameter indicates that the ssDNA behaves as the Zimm polymer. Also the dsDNA can be identified as being predominantly of the Zimm type. For long flexible polymers, the short-time behavior of the MSD of the polymer end is usually (particularly, in Ref. [3]) described by power laws $\sim t^{1/2}$ in the Rouse limit and $\sim t^{2/3}$ in the Zimm case. However, the applicability of these laws is very limited and, as discussed in our paper [1], their use in the interpretation of the experiment [3] is flawed. This conclusion has been recently supported by the fluorescence correlation spectroscopy study [6], according to which the Rouse-type behavior reported for dsDNA in [3] has been clearly ruled out.

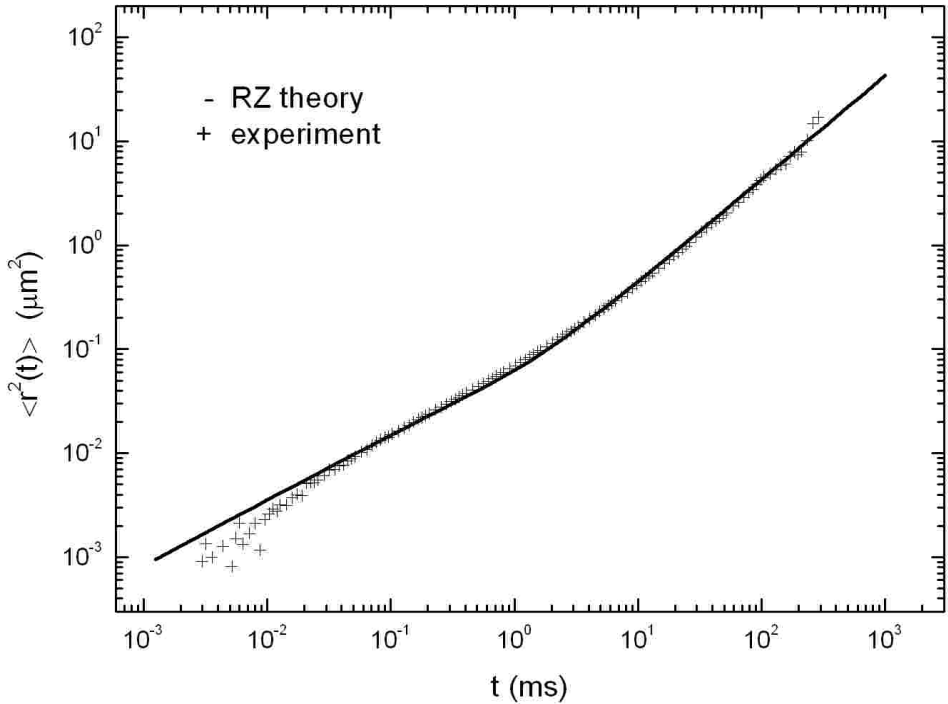

**Fig. 2.** MSD for the Rouse-Zimm model optimized to the experimental data [4] for ssDNA (6700 bases) at $T = 310$ K and $\eta = 0.69$ mPa s. The optimization yielded the polymer parameters $a = 8.03$ nm, $b = 3.94$ nm, and $N = 474$. The draining parameter is $h \approx 20.88$.

*Acknowledgments*. We are greatly indebted to O. Krichevsky for providing us with the experimental data [4] prior to publication. This work was supported by the grants VEGA 1/3033/06 and 1/4021/07 from the Scientific Grant Agency of the Slovak Republic.